\documentclass[twocolumn,showpacs,preprintnumbers,amsmath,amssymb]{revtex4}

\usepackage{graphicx}
\usepackage{dcolumn}
\usepackage{bm}

\begin{document}


\title {Shape transformations of a compartmentalized fluid surface}

\author{Hiroshi Koibuchi}
 \email{koibuchi@mech.ibaraki-ct.ac.jp}
\affiliation{%
Department of Mechanical and Systems Engineering, Ibaraki National College of Technology, Nakane 866 Hitachinaka, Ibaraki 312-8508, Japan}%

\date{\today}

\begin{abstract}
A surface model on compartmentalized spheres is studied by using the Monte Carlo simulation technique with dynamical triangulations. We found that the model exhibits a variety of phases:  the spherical phase, the tubular phase, the planar phase, the wormlike planar phase, the wormlike long phase, the wormlike short phase, and the collapsed phase. It is also shown that almost all phases are separated from their neighboring phases by first-order transitions. Mechanical strength of the surface is given only by elastic skeletons, which are the compartment boundaries, and vertices diffuse freely inside the compartments. We confirm that the cytoskeletal structure and the lateral diffusion of vertices are an origin of such a variety of phases.

\end{abstract}

\maketitle
\section{Introduction}\label{intro}
Biological membranes and synthetic polymer-membranes show a variety of shapes such as spherical, tubular, discoid, cylindrical, and many others including starfish \cite{SEIFERT-LECTURE2004}. The shape of membranes is partly understood numerically with a surface model called a minimal model \cite{SBR-PRA1991} and also with the area difference bilayer model \cite{EVANS-BPJ1974,JSWW-PRW1995}. External forces such as gravity and flow fields make the surface shape change \cite{KSL-EPL1995,KWSL-PRL1996,NG-PRL2004}.

The surface shape can also be influenced by thermal fluctuations \cite{DS-EPL1996,PM-EPL2002,GK-SMMS2004}. Therefore, the membrane shape should be understood as an equilibrium statistical mechanical phenomenon, although the shape of membranes seems to have a non-equilibrium nature. We should remind ourselves that the shape transformation and the surface fluctuation are two different phenomena, where the surface fluctuation phenomena have been extensively studied statistical mechanically \cite{NELSON-SMMS2004-1,Bowick-PREP2001,Gompper-Schick-PTC-1994,Peliti-Leibler-PRL1985,DavidGuitter-EPL1988,PKN-PRL1988}.  

The surface fluctuation transition is accompanied by the collapsing transition in artificial membranes \cite{CNE-PRL-2006} and in the surface model \cite{KD-PRE2002,KOIB-PRE-20045-NPB-2006} of Helfrich \cite{HELFRICH-1973}, Polyakov \cite{POLYAKOV-NPB1986}, and Kleinert \cite{KLEINERT-PLB1986}. This also indicates that the shape of membranes should be understood within the context of the theory of phase transitions. In fact, shape transformations, such as the prolate-oblate transition driven by the thermal fluctuation, were experimentally observed \cite{DS-EPL1996}. Current understanding of the effects of thermal fluctuations on the conformation and the elastic properties of membranes are reviewed in \cite{GK-SMMS2004}. In \cite{KOIB-PRE2007-2}, it was shown that the shape of a compartmentalized fluid surface model changes due to thermal fluctuations. The results suggest that possible origins for the variety of membrane shapes are the cytoskeletal structure and the fluidity of lipids in membranes. Moreover, it was also suggested in \cite{KOIB-PRE2007-2} that the large variety of shapes can be understood in the framework of a surface model which has a cytoskeleton. The cytoskeletal structure has been considered as a key notion for understanding physics of membranes \cite{MSWD-PRE-1994,HHBRM-PRL-2001,Kusumi-BioJ-2004}. 

In this article, in order to make this observation more convincing we study another compartmentalized fluid surface model, which is almost identical to the model in \cite{KOIB-PRE2007-2}. It is remarkable that a small change in the model makes a large difference in the multitude of surface shapes. The only difference between the model in this article and that of Ref.\cite{KOIB-PRE2007-2} is in the junction elasticity; rigid plates are assumed as the junctions in \cite{KOIB-PRE2007-2} while neither two-dimensional elasticity nor rigid plate is assumed in the model of this article. Both of the compartmentalized models are inhomogeneous because of the cytoskeletal structures; the surface strength on the compartment boundary is different from that inside the compartments, and moreover the diffusion of vertices is confined only in the compartments. 
\section{Model}\label{model}
The compartmentalized structure is a sublattice on a triangulated surface, which is constructed from the icosahedron. By dividing the edges of the icosahedron into $\ell$ pieces, we have a triangulated lattice of size $N\!=\!10\ell^2+2$, which is the total number of vertices. Then, we have a sublattice of size $N_S\!=\!30m\ell$ in the $N\!=\!10\ell^2\!+\!2$ lattice if $m$ divides $\ell$. The vertices in the sublattice include the junctions of the compartments on the $N\!=\!10\ell^2\!+\!2$ lattice, and the total number of junctions $N_J$ is given by $N_J\!=\!10m^2\!+\!2$. The total number of links between the junctions is $3N_J\!-\!6$, and each link contains $\ell/m$ vertices. Thus we have $N_S\!=\!30m\ell$. The compartment size can be characterized by $n\!=\!\sum_{i=1}^{(\ell/m)-2} i$, which is the total number of vertices in a compartment.

\begin{figure}[htb]

\unitlength 0.1in
\begin{picture}( 0,0)(  10,10)
\put(12,9.0){\makebox(0,0){(a) }}%
\put(23,9.0){\makebox(0,0){(b) }}%
\put(33,9.0){\makebox(0,0){(c) }}%
\end{picture}%
\includegraphics[width=2.5cm]{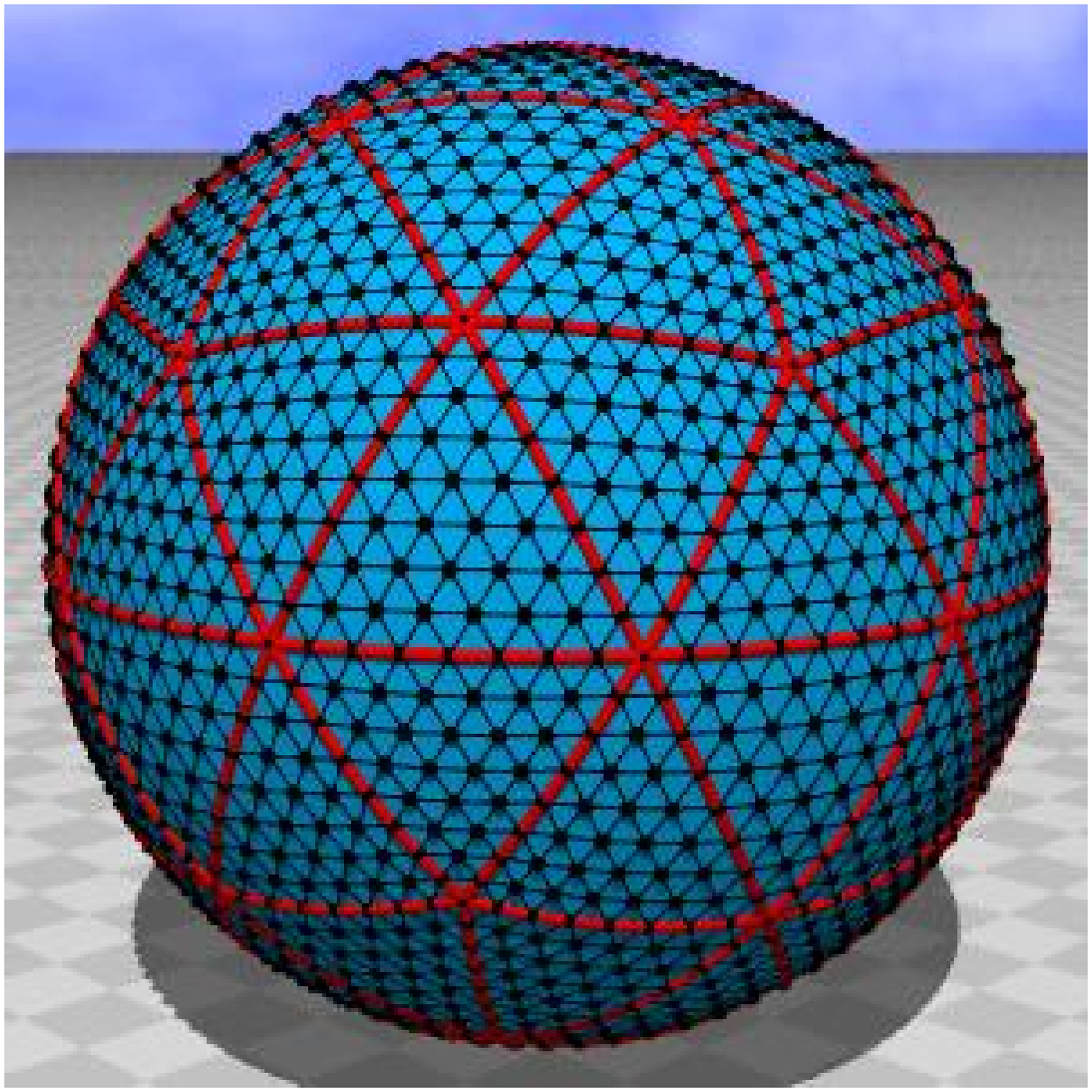}
\includegraphics[width=5.0cm]{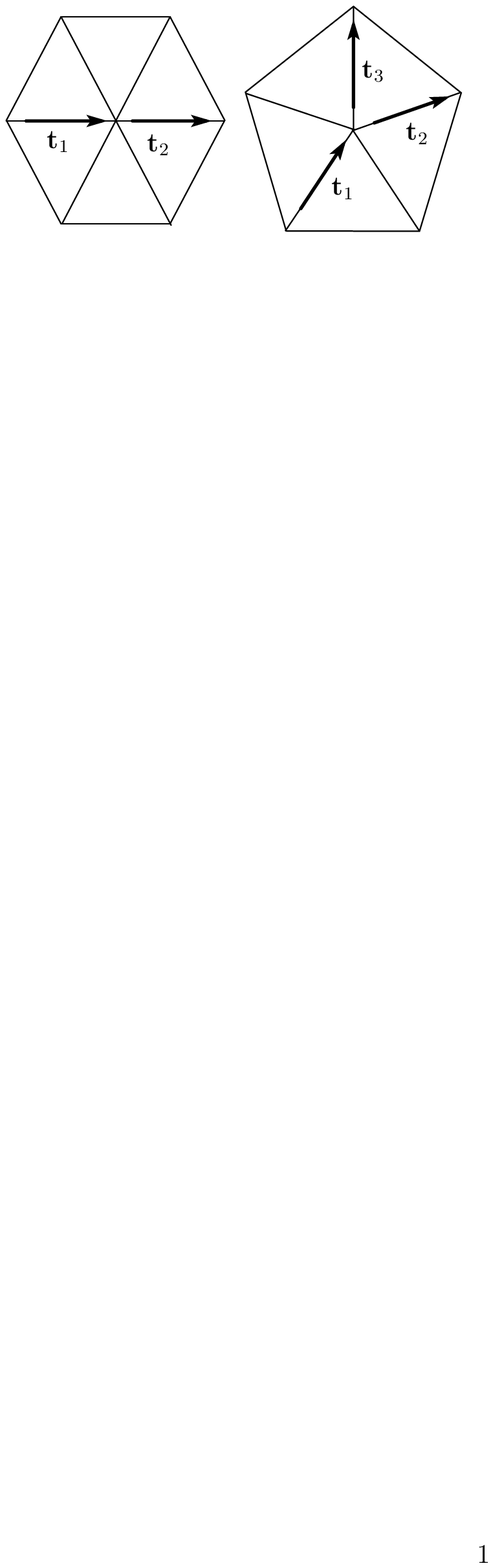}

\caption{(Color online) (a) Starting configuration of surfaces of size $(N,N_S,N_J)\!=\!(2562,960,42)$; thick lines denote a sublattice composed of the linear chains and the junctions, (b) tangent vectors  ${\bf t}_1$ and ${\bf t}_2$ at a vertex with coordination number $q\!=\!6$ that give rise to a contribution $1\!-\!{\bf t}_1 \cdot {\bf t}_2$ to the  bending energy with the weight of $1$, and (c) tangent vectors  ${\bf t}_1$, ${\bf t}_2$ and ${\bf t}_3$ at a vertex with coordination number $q\!=\!5$ that contribute $1\!-\![{\bf t}_1 \cdot ({\bf t}_2\!+\!{\bf t}_3 )]/2$ to the bending energy with the weight of $1/2$. }
\label{fig-1}
\end{figure}

Figure \ref{fig-1}(a) shows the starting configuration for Monte Carlo simulations. The size of surface is characterized by two integers $(\ell,m)\!=\!(16,2)$, and the size is given by  $(N,N_S,N_J)\!=\!(2562,960,42)$, and $n\!=\!21$. 

The model is defined by the Gaussian bond potential $S_1$ and the one-dimensional bending energy $S_2$, which are respectively defined by
\begin{equation}
\label{Disc-Eneg} 
S_1=\sum_{(ij)} \left(X_i-X_j\right)^2,\quad S_2=\sum_{(ij)} \left( 1-{\bf t}_i \cdot {\bf t}_j \right),
\end{equation} 
where $X_i$ is the three-dimensional position of the vertex $i$ and ${\bf t}_i$ is a unit tangent vector of the bond $i$. $\sum_{(ij)}$ in $S_1$ is the sum over all bonds $(ij)$ on the lattice, and $\sum_{(ij)}$ in $S_2$ is the sum over all nearest neighbor bonds $(ij)$ on the sublattice. 

Tangent vectors at the junctions of coordination numbers $q\!=\!6$ and  $q\!=\!5$ are shown in Figs.\ref{fig-1}(b) and \ref{fig-1}(c). The tangent vectors ${\bf t}_1$ and ${\bf t}_2$ at a junction of coordination number $q\!=\!6$ give rise to a contribution $1\!-\!{\bf t}_1 \cdot {\bf t}_2$ to $S_2$ with the weight of $1$. The remaining two inner-products of tangent vectors are defined just like $1\!-\!{\bf t}_1 \cdot {\bf t}_2$ at the $q\!=\!6$ vertices. On the contrary, the tangent vectors ${\bf t}_1$, ${\bf t}_2$ and  ${\bf t}_3$ at a junction with coordination number $q\!=\!5$ contribute $1\!-\!\left[{\bf t}_1 \cdot ({\bf t}_2\!+\!{\bf t}_3)\right]/2$ to the bending energy with the weight of $1/2$. The remaining four inner-products of tangent vectors are defined just like $1\!-\!\left[{\bf t}_1 \cdot ({\bf t}_2\!+\!{\bf t}_3)\right]/2$ at the $q\!=\!5$ vertices; this is the reason for the weight $1/2$ of the bending energy $S_2$ at the $q\!=\!5$ junctions. Consequently, the definition of the bending energy at the $q\!=\!6$ junctions is almost identical to that at the $q\!=\!5$ junctions, whose total number is only $12$.

The partition function $Z$ of the model is given by
\begin{eqnarray} 
\label{Part-Func}
 Z = \sum_{\cal T} \int^\prime \prod _{i=1}^{N} d X_i \exp\left[-S(X,{\cal T})\right],\\  
 S(X,{\cal T})=S_1 + b S_2, \nonumber
\end{eqnarray} 
where $S(X,{\cal T})$ is the Hamiltonian, and $b[kT]$ is the bending rigidity, which is a microscopic quantity and therefore, it is not always identical to the macroscopic bending rigidity. $\sum_{\cal T}$ denotes all possible triangulations which keep the compartment boundary (= the sublattice bonds) unchanged.  $\int^\prime\prod _{i=1}^{N} d X_i$ denotes the multiple three-dimensional integrations under the condition that the center of mass of the surface is fixed.

\section{Monte Carlo technique}\label{MC-Techniques}
The integrations of the dynamical variables $X$ and ${\cal T}$ are performed by the Monte Carlo simulation technique \cite{KANTOR-NELSON-PRA1987,WHEATER-NPB1996,Baum-Ho-PRA1990,CATTERALL-PLB1989,AMBJORN-NPB1993}. The three dimensional random shift $\delta X$ of $X$ generates a new position $X^\prime\!=\!X\!+\!X$, which is accepted with the probability ${\rm Min} [1, \exp(-\Delta S)]$, where $\Delta S\!=\! S({\rm new})\!-\!S({\rm old})$. The vertices can be classified into three groups; the vertices inside the compartment, the vertices on the compartment boundary, and the vertices at the junctions; the final two groups of vertices are those constructing the sublattice. The point $\delta X$ is randomly chosen in a sphere and, the radius of the sphere is fixed at the beginning of the simulations so that the acceptance rate is equal to about $50\%$ in each group of vertices. The radius assumed for one group of vertices is not always identical to those for the other groups of vertices. The summation over ${\cal T}$ is performed by the standard bond flip technique \cite{Baum-Ho-PRA1990,CATTERALL-PLB1989,AMBJORN-NPB1993} and, therefore the acceptance rate for the flip is not fixed a priori but found to vary approximate bracket $70\%\sim 75\%$, which is slightly dependent on $b$. We assume the surfaces of size $(N,N_S,N_J)\!=\!(5762,2160,92)$ and $(N,N_S,N_J)\!=\!(10242,3840,162)$, which correspond to integers $(\ell,m)\!=\!(24,3)$ and $(\ell,m)\!=\!(32,4)$. The total number of MCS (Monte Carlo sweep) after the thermalizaion is about $1\!\times \!10^8\sim 1.5\!\times \!10^8$ for the $N\!=\!5762$ surface and $1.3\!\times \!10^8\sim 2\!\times \!10^8$ for the $N\!=\!10242$ surface. The thermalization process comprises about $1\!\times \!10^8$ MCS, which is sufficiently large, in almost all cases.  

\section{Results of simulation}\label{Results}
\begin{figure}[htb]
\includegraphics[width=8.5cm]{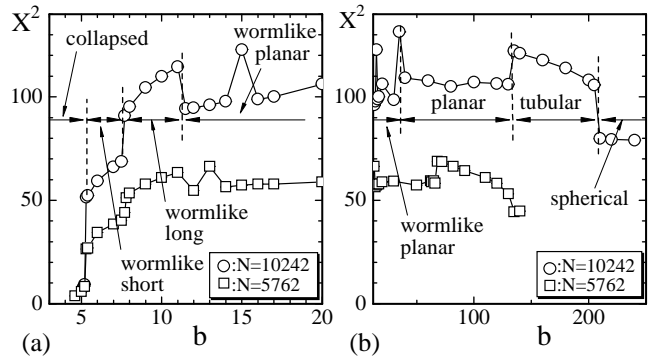}  
\caption{The mean square size $X^2$ versus $b$ at (a) small $b$ region $b\!\leq\!20$ and at (b) whole region $b\!\leq\!240$. The compartment size is given by $n\!=\!21$, which is the total number of vertices in a compartment. } 
\label{fig-2}
\end{figure}
The shape of surfaces can be reflected in the mean square size $X^2$, which is defined by
\begin{equation}
\label{Mean-Square-Size}
X^2= {1\over N} \sum _i \left( X_i-\bar X \right)^2,\quad \bar X = {1\over N} \sum_i X_i,
\end{equation}
where $\bar X$ is the center of the surface. Figures \ref{fig-2}(a) and \ref{fig-2}(b) show $X^2$ versus $b$ obtained at relatively small $b$ region and at whole $b$ region, respectively. The vertical dashed lines denote phase boundaries where $X^2$ discontinuously changes. We have at least seven phases in the region $0<b<240$ on the surface of size $(N,N_S,N_J)\!=\!(10242,3840,162)$. We call the phases as the collapsed, the wormlike short, the wormlike, the wormlike planar, the planar, the tubular, and the spherical. Almost all of two neighboring phases, except the wormlike planar phase and the planar phase, seem to be connected by a first-order transition, because $X^2$ discontinuously changes at the phase boundaries. 

$X^2$ in the wormlike planar phase of the $N\!=\!10242$ surface is wildly fluctuating on the $b$ axis; $X^2$ at $b\!=\!15$ and those at the phase boundary close to the planar phase are different from the remaining $X^2$ in the wormlike planar phase. $X^2$ at $b\!=\!13$ of the $N\!=\!5762$ surface also seems to be an anomalous value. The surface shape at $b\!=\!15$ of the $N\!=\!10242$ surface and that at $b\!=\!13$ of the $N\!=\!5762$ surface are wormlike, and then we understand that the configuration was trapped in the potential minimum corresponding to the wormlike long phase in the simulations. The potential barriers separating the phases seem to be low because the surface size is not sufficiently large, and for this reason such anomalous behavior of $X^2$ can be seen in the wormlike planar phase.  We must emphasize that the anomalous behavior of $X^2$ does not imply that the model is ill-defined. In fact, the Hamiltonian such as the bending energy $S_2$ is not unstable and has the unique value corresponding to the given value of $b$ even in the wormlike planar phase as we will see later in this paper.

\begin{figure*}[htb]
\unitlength 0.1in
\begin{picture}( 0,0)(  10,10)
\put(15,9.0){\makebox(0,0){(a) $b\!=\!5.3$ }}%
\put(27,9.0){\makebox(0,0){(b) $b\!=\!10$ }}%
\put(40,9.0){\makebox(0,0){(c) $b\!=\!100$ }}%
\put(53,9.0){\makebox(0,0){(d) $b\!=\!205$ }}%
\put(66,9.0){\makebox(0,0){(e) $b\!=\!210$ }}%
\end{picture}%
\centering
\includegraphics[width=16.0cm]{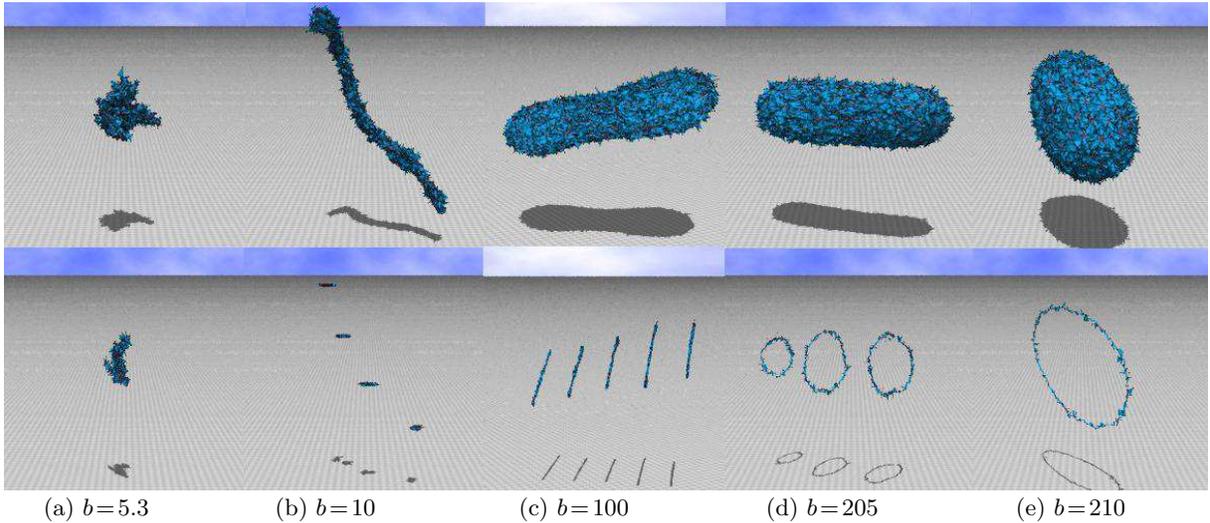}
\caption{(Color online) The snapshots of surfaces and the surface sections of size $(N,N_S,N_J)\!=\!(10242,3840,162)$ obtained at (a) $b\!=\!5.3$ (collapsed phase), (b) $b\!=\!10$ (wormlike long phase), (c) $b\!=\!100$ (planar phase), (d) $b\!=\!205$ (tubular phase), and  (e) $b\!=\!210$ (spherical phase). } 
\label{fig-3}
\end{figure*}
Snapshots of surfaces and their sections are shown in Figs.\ref{fig-3}(a)--\ref{fig-3}(e), which were respectively obtained in the collapsed phase, the wormlike long phase, the planar phase, the tubular phase, and the spherical phase. The surfaces and the surface sections were shown in the same scale.  The self-avoiding property \cite{GREST-JPIF1991,BOWICK-TRAVESSET-EPJE2001,BCTT-PRL2001} is not assumed in our model and, therefore the phase structure in the small $b$ region seems phantom. However, as we see in the snapshots, the phase structure seems realistic in the large $b$ region.

\begin{figure*}[htb]
\unitlength 0.1in
\begin{picture}( 0,0)(  10,10)
\put(14,9.0){\makebox(0,0){(a) $b\!=\!6$ }}%
\put(30,9.0){\makebox(0,0){(b) $b\!=\!10$ }}%
\put(46,9.0){\makebox(0,0){(c) $b\!=\!14$ }}%
\put(62,9.0){\makebox(0,0){(d) $b\!=\!35$ }}%
\end{picture}%
\centering
\includegraphics[width=16.0cm]{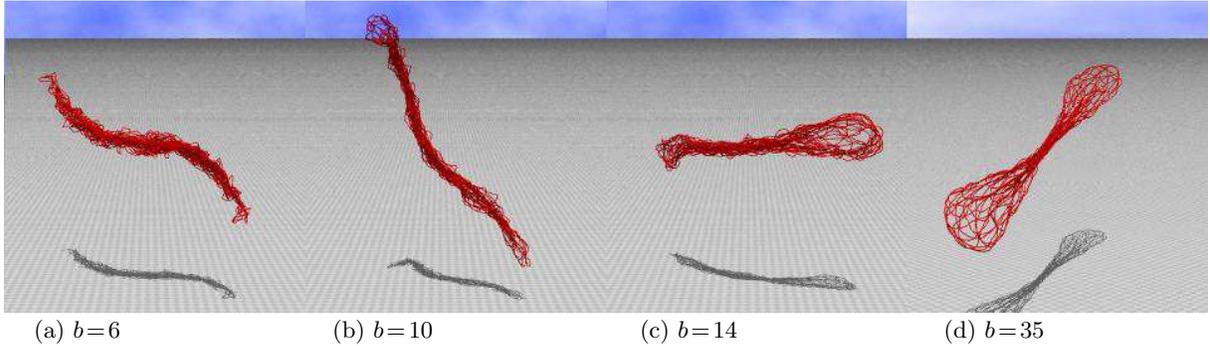}
\caption{(Color online) The snapshots of skeletons obtained at (a) $b\!=\!6$ (wormlike short phase), (b) $b\!=\!10$ (wormlike long phase), (c) $b\!=\!14$ (wormlike planar phase), and  (d) $b\!=\!35$ (wormlike planar phase). The surface size is given by $(N,N_S,N_J)\!=\!(10242,3840,162)$.} 
\label{fig-4}
\end{figure*}
In order see the difference between the wormlike short phase, the wormlike long phase, and the wormlike planar phase, we show snapshots of skeletons in Figs.\ref{fig-4}(a)--\ref{fig-4}(d), which were obtained in the wormlike short phase (Fig.\ref{fig-4}(a)), the wormlike long phase (Fig.\ref{fig-4}(b)), and the wormlike planar phase (Figs.\ref{fig-4}(c) and \ref{fig-4}(d)). All figures were drawn in the same scale. We understand from the snapshots in Figs.\ref{fig-4}(c) and \ref{fig-4}(d) that one part of the surface is wormlike and the remaining part is planar in the wormlike planar surfaces. We see from Figs.\ref{fig-4}(c) and \ref{fig-4}(d) that the size of planar part varies depending on $b$ in the wormlike planar phase; we see the inflated parts are planar from their surface sections. It is easy to understand that $X^2$ is strongly dependent on the size of the planar part. $X^2$ also depends on the number of planar parts; we see two planar parts at the two ends of the surface in Fig.\ref{fig-4}(d). For this reason, $X^2$ wildly fluctuates at the phase boundary ($b\!\simeq\!35$) between the wormlike planar phase and the planar phase as mentioned above. 

The surface shape in the wormlike short phase is also wormlike as we see in Fig.\ref{fig-4}(a), however, the thickness or equivalently the longitudinal length of surface is slightly different from those of the surfaces in the wormlike long phase. This difference is reflected in $X^2$, and consequently the wormlike short phase is separated from the wormlike long phase by the first-order transition.

\begin{figure}[htb]
\centering
\includegraphics[width=8.5cm]{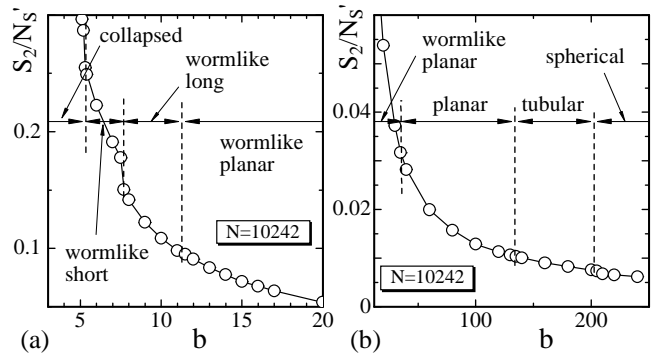}  
\caption{The bending energy $S_2/N_S^\prime$ versus $b$ at (a) small $b$ region $b\!\leq\!20$ and at (b) whole region $b\!\leq\!240$. } 
\label{fig-5}
\end{figure}
The one-dimensional bending energy $S_2/N_S^\prime$ versus $b$ is shown in Figs.\ref{fig-5}(a) and \ref{fig-5}(b), where $N_S^\prime$ is the total number of vertices where $S_2$ is defined. The junctions of coordination number $q\!=\!6$ ($q\!=\!5$) are counted $3$ ($2.5$) times in $N_S^\prime$ because of the definition of $S_2$ and, therefore $N_S^\prime$ is given by $N_S^\prime\!=\!N_S\!+\!2N_J\!-\!6$, which is also written by $N_S^\prime\!=\!10\ell^2\!-\!60m^2\!+\!2$. The vertical dashed lines in the figures denote the phase boundaries. We find a discontinuous change in $S_2/N_S^\prime$ at the boundaries between the collapsed phase and the wormlike short phase and at the boundary between the wormlike short phase and the wormlike long phase. No discontinuous change can be seen in $S_2/N_S^\prime$ at any other phase boundaries. 

We see no wild fluctuation of $S_2/N_S^\prime$ in the wormlike planar phase in Figs.\ref{fig-5}(a) and \ref{fig-5}(b). $S_2/N_S^\prime$ smoothly vary even at $b\!=\!15$ and at $b\!\simeq\!35$, where $X^2$ wildly fluctuates.

\begin{figure}[htb]
\centering
\includegraphics[width=8.5cm]{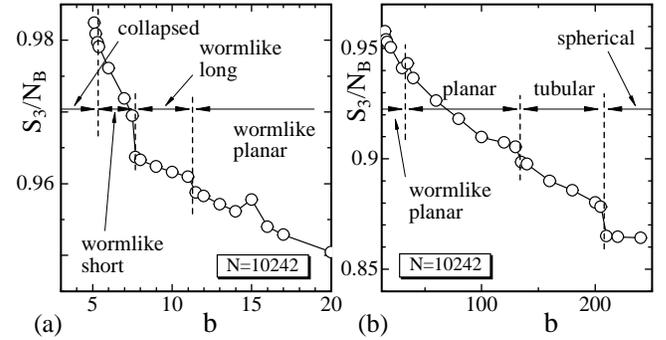}  
\caption{The two-dimensional bending energy $S_3/N_B$  versus $b$ at (a) small $b$ region $b\!\leq\!20$ and at (b) whole region $b\!\leq\!240$.} 
\label{fig-6}
\end{figure}
The two-dimensional bending energy is defined by
\begin{equation}
\label{two-dim-bending}
S_3=\sum_{(ij)}\left(1-{\bf n}_i\cdot {\bf n}_j\right), 
\end{equation}
where ${\bf n}_i$ is a unit normal vector of the triangle $i$. The surface fluctuation can be reflected in $S_3$, which is not included in the Hamiltonian. Figures \ref{fig-6}(a) and \ref{fig-6}(b) show $S_3/N_B$ versus $b$, where $N_B\!=\!3N\!-\!6$ is the total number of bonds. Discontinuous changes can be seen in $S_3/N_B$ at the boundary between the wormlike short phase and the wormlike long phase, at the boundary between the wormlike long phase and the wormlike planar phase, at the boundary between the planar phase and the tubular phase, and at the boundary between the tubular phase and the spherical phase. The discontinuous changes in $S_3/N_B$ are consistent to those in $X^2$ in Figs.\ref{fig-2}(a) and \ref{fig-2}(b). Note also that anomalous spikes of $S_3/N_B$ at $b\!=\!15$ and at $b\!=\!35$ correspond to the anomalous value or the wild fluctuations of $X^2$ mentioned above. 

We have seen that at least one physical quantity discontinuously changes at the phase boundaries except the boundary between the planar phase and the wormlike planar phase. At this boundary we see that no physical quantity discontinuously changes although $X^2$ anomalously fluctuated, which was seen in Fig. \ref{fig-1}(b). Then, the discontinuous nature of the transition at this boundary is not confirmed from the numerical data in this paper. Therefore, we consider that almost all phases, except the planar and the wormlike planar phases, are separated from their neighboring phases by first-order transitions.

\begin{figure}[htb]
\centering
\includegraphics[width=8.5cm]{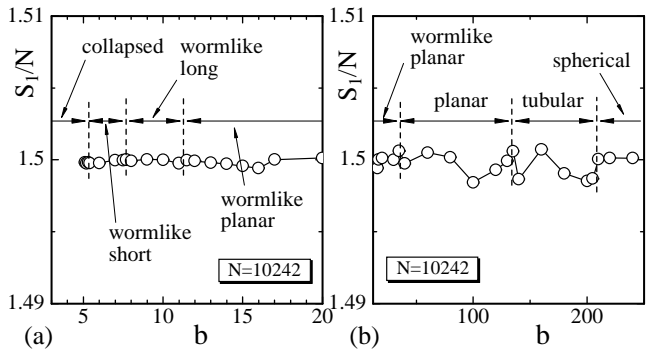}  
\caption{The Gaussian bond potential $S_1/N$ versus $b$ at (a) small $b$ region $b\!\leq\!20$ and at (b) whole region $b\!\leq\!240$.} 
\label{fig-7}
\end{figure}
The Gaussian bond potential $S_1/N$ is expected to be $S_1/N\!\simeq\!3/2$ because of the scale invariance of the partition function. Figures \ref{fig-7}(a) and \ref{fig-7}(b) show that the expected relation is almost satisfied. We find that the relation is satisfied in the region of low bending rigidity, where the surfaces are almost collapsing. On the contrary, we find that the relation is not exactly satisfied in the region of high bending rigidity, where the surfaces are inflated, although the deviation is very small compared to the value itself.

\section{Summary and conclusion}\label{Conclusions} 
To summarize the results, we have investigated a compartmentalized fluid surface model by using the canonical MC simulation technique and found a variety of phases; the spherical phase, the tubular phase, the planar phase, the wormlike planar phase, the wormlike long phase, the wormlike short phase, and the collapsed phase. Almost all two neighboring phases are connected by first-order transitions. The spherical phase and the tubular phase are connected by a first-order transition, which is quite similar to the prolate-oblate transition. Our results indicate that the variety of membrane shapes and their transformations can be understood in the inhomogeneous model, which is characterized by compartmentalization of fluidity of vertices and the cytoskeletal structure constructed on the conventional homogeneous surface model. 

It is interesting to study the model by including the two-dimensional bending energy in the Hamiltonian. The phase structure in the thermodynamic limit and the dependence of the phase structure on the compartment size still remains to be clarified.

This work is supported in part by a Grant-in-Aid for Scientific Research from Japan Society for the Promotion of Science.  



\end{document}